\documentclass[twocolumn,showpacs,preprintnumbers,amsmath,amssymb,eps]{revtex4}
\usepackage{epsfig}
\usepackage{graphicx}
\usepackage{dcolumn}
\usepackage{bm}

\newcommand{\be}{\begin{equation}} \newcommand{\ee}{\end{equation}}
\newcommand{\beqn}{\begin{eqnarray}} \newcommand{\eqn}{\end{eqnarray}}
\newcommand{\nb}{\nonumber}
\newcommand{\lb}{\label}

\begin{document}

\title{Magnetic Properties  Calculations of Cuprate Superconductors
based on a Phase Separation Theory}
\author{D. N. Dias$^{\rm 1}$}
\author{E. S. Caixeiro$^{\rm 2}$}
\author{E. V. L. de Mello$^{\rm 1}$}
\affiliation{$^{\rm 1}$%
Instituto de F\'{\i}sica, Universidade Federal Fluminense, Niter\'oi, RJ 24210-340, Brazil\\}%
\affiliation{$^{\rm 2}$Centro Brasileiro de Pesquisas F\'{\i}sicas, Rio de
Janeiro, RJ 22290-180 Brazil\\}
\date{\today}
\begin{abstract}

There is a great debate concerning the hole of the inhomogeneities
in high critical temperature superconductors (HTS). In this context,
there are many experiments and proposals related with a possible
electronic phase separation (PS). However there is not a method to
quantify how such transition occurs and how it develops. The
Cahn-Hilliard (CH) theory of phase separation provides a way which
we can trace the phase separation process as a function of
temperature in agreement with some experiments. Here we coupled
these calculations, with parameters that yield a stripe like
pattern, to the Bogoliubov-deGennes (BdG) approach to an
inhomogeneous superconductor in order to derive many HTS properties
of the $La_{2-x}SrCuO_4$ (LSCO) system. Taking the upper pseudogap
as the PS transition line, we can show that; the onset of
superconductivity follows close the Nernst signal, the leading edge
shift is close to the zero temperature average gap and the
superconducting phase is achieved by percolation or Josephson
coupling. Our approach is also suitable to reproduce  the
experimental measurements of the $H_{c2}$ field and explain why it
does not vanish above $T_c$.

\end{abstract}


\maketitle

\section{Introduction}

The measurement of the pseudogap and many nonconventional properties
in the normal phase\cite{TS,Tallon} of HTS has become a long
standing puzzle. It is a general consensus that understanding these
properties is crucial to comprehend the nature of the
superconducting transition and the fundamental interaction in these
materials.
It is  quite interesting  that while low critical temperature
superconductors have a well characterized normal phase, the nature
of the normal or pseudogap  phase of HTS is an open problem which
remains a matter of intense experimental and theoretical study\cite{Lee,Moura}.

One of the possible reason why this problem remains unsolved after
20 years of intense research may be due to the uncontrolled
intrinsic inhomogeneities in some HTS, which as some of us have
argued\cite{Mello04}, may also depend on the sample preparation
method. Since the theoretical
prediction\cite{Zaanen} and the detection of
stripes\cite{Tranquada}, it is clear that at least some family of
compounds exhibit some degree of non-uniformities. In fact, the
role of the inhomogeneities is an open question: it seems not important
in some experiments\cite{Bobroff,Loram}, but on the other hand, the detection and
the effect of inhomogeneities have rendered many articles and
books\cite{Muller,Trieste}, as we will discuss in detail in
this paper. 

These unusual features of cuprates, like the stripes, led
to theoretical proposals that PS is essential to understand their
physics\cite{Zaanen,Machida,EK,Carlson}. Indeed PS has been observed
on the $La_2CuO_{4+\delta}$ by x-ray and transport
measurements\cite{Grenier,Jorg}, which have detected a spinodal
phase segregation into an oxygen-rich (or hole-rich) metallic phase
and an oxygen-poor antiferromagnetic (AF) phase above T=220K. There
is also evidence of ion diffusion at room temperature in micro
crystals of the Bi2212 superconductors at a very slow
rate\cite{Truccato}.

Recent angle resolved photoemission (ARPES) experiments with
improved energy and momentum  resolution\cite{Harris,DHS,Zhou,Zhou04,Ino}
have distinguished two  electronic components in $\vec k$-space
associated with the $La_{2-x}Sr_xCuO_4$ (LSCO) system: a metallic
quasi particle spectral weight at the ($\pi/2,\pi/2$) nodal
direction which increases with hole doping and an insulator like
spectral weight at the  end of the Brillouin zone straight segments
in the ($\pi,0$) and (0,$\pi$) antinodal regions which are almost
insensitive to the doping level.

New STM data with great resolution have also revealed strong
inhomogeneities in the form of a patchwork of (nanoscale) local
spatial variations in the density of states which, at low
temperature, is related to the local superconducting
gap\cite{Fournier,Pan,Davis,McElroy}. More recently it was possible
to distinguish two distinct behavior: well defined coherent and
ill-defined incoherent peaks depending on the exactly spectra
location at a $Bi_2Sr2CaCu_2O_{8+\delta}$ (Bi2212)
surface\cite{Hoffman,McElroy1}. STM experiments have also detected a
regular low energy checkerboard order in the electronic structure of
the Bi2212 family at low temperature\cite{McElroy2}, above the
superconducting critical temperature ($T_c$)\cite{Vershinin} and in
the $Na_xCa_{2-x}CuO_2Cl_2$\cite{Hanaguri}.

Bulk sensitive experiments like nuclear magnetic and quadrupolar
resonance (NMR and NQR) have also provided ample evidence for
spatial charge inhomogeneity in the $CuO_2$ of some HTS
planes\cite{Curro,Singer,Haase}. Singer at al\cite{Singer} measured
a distribution of $T_1$ over the Cu NQR spectrum in bulk LSCO which
can be attributed to a  distribution of holes $p$ with a half width
of $\Delta p/p\approx 0.5$. The increasing of $\Delta p/p$ as the
temperature decreases is likely {\it  the strongest indication of a
PS transition} at temperatures above 600K in HTS. More recently, NMR
results on $La_{1.8-x}Eu_{0.2}SrCuO_{4}$ were interpreted as
evidence for a spatially inhomogeneous charge distribution in a
system which the spin fluctuations are suppressed\cite{NCurro}. This
new result is also a clear indication that the charge disorder may
be due to a phase separation transition.

In this paper we work out in detail the scenario to the physics of
HTS based on a PS transition at the (upper) pseudogap curve given in the
review of Tallon at al\cite{Tallon}, which starts at very high
temperatures ($\approx 1000$K) in the low underdoped region and
falls to zero near the average doping level $\rho_m =0.2$. This PS
transition is treated by the CH theory\cite{CH}, originally proposed
to describe the PS transition in alloys, and yields two equilibria
densities; one low and other with high values which grows apart as
the temperature is lowered, exactly as seen in the NQR
experiments\cite{Singer} or, indirectly by the stripes structures.
Applying the  BdG theory of superconductivity to such a disordered
medium we see that, as the temperature decreases, the
superconductivity appears in nanoscale regions inside the high density
phase.
With this approach, we can show that:\\
i- the zero temperature average superconducting gap and the
amplitude of pairing $|\Delta|$ as function of the average doping
level $\rho_m$ is in reasonable agreement with the superconducting
state gap\cite{Harris} and the
leading edge shift measured by ARPES\cite{Ino}.\\
ii-The onset temperature of superconductivity for each compound
is close the lower pseudogap temperature\cite{TS,Moura,Mello04}
and also the recent measurements on the onset temperature of Nernst
signal\cite{Nernst1,Nernst2,Nernst3}, providing a
simple interpretation to these experiments.\\
iii-The calculation of $H_{c2}(T)$ for a system of non-uniform
density regions with different local superconducting $T_c(i)$
($T_c(i)$ will be defined below) is in excellent agreement with the
measurements in the LSCO system and also provides an explanation why
the $H_{c2}(T)$ field does not vanish at $T_c$.

Our proposal is an alternative scenario to the phase-disordered
theory\cite{EK,Carlson} which has gained increased
attention\cite{Nernst1,Nernst2,Nernst3,M2S,Ong,Alloul}. Contrary to the
familiar BCS theory in which the complex superconducting order
parameter $\Psi=|\Delta|e^{i\theta}$ develops with the phase
$\theta$ essentially locked, in their calculations\cite{EK,Carlson}
thermally generated vortices destroy long range phase coherence at
temperatures close to the superconducting critical temperature
$T_c$. The temperature $T_{\theta}$ which phase rigidity is lost was
estimated to be very low in the underdoped region and to increase
continuously with doping\cite{EK,Carlson}. Therefore, in this
scenario, the pseudogap phase is characterized by the presence of
Cooper pairs  with nonvanishing  pairing amplitude $|\Delta|$ but,
due to thermally excited vortices (in zero field), without phase
rigidity.

The presence of the pairing amplitude $|\Delta|$ has been clearly
detected by many spectroscopy experiments starting at a temperature
that we call $T^*$, the lower pseudogap temperature, 
which,  depending on the compound, can vary
from $T^*\approx T_c$ to roughly 100K above
$T_c$\cite{Moura,Loeser,Ding,Renner}. On the other hand, Nernst
effect experiments which measure a voltage transverse to a thermal
gradient in the presence of a magnetic field perpendicular to the
superconducting film are specially sensitive to the existence and
the drift of vortices\cite{Nernst1,Nernst2,Nernst3}. Consequently a
large Nernst signal above $T_c$, which starts at $T_{onset} \approx T^*$,
 was taken as a strong indication of
a vortex-like behavior and the presence of a large region of
fluctuating superconductivity\cite{Nernst1,Nernst2,Nernst3,M2S}.
Thus the important question is whether this fluctuating region is
due to the thermally induced vortices of the phase disordered
scenario or to some other mechanism, like a non-uniform density.

Recently, in order to address this question and to study the
pseudogap region, a series of combined measurements on the Nernst
effect, the upper critical field $H_{c2}$ and magnetization above
the $T_c$ on different compounds were performed by Wang et
al\cite{Ong}. The results were interpreted as providing strong
support to the phase-disordered theory\cite{Ong}. On the other hand,
Nernst effect studies carried on the presence of induced disordered
demonstrated that $T_{onset}$ remained basically the same but $T_c$
decreased considerably with the presence of controlled
defects\cite{Alloul} and they have detected fluctuations only near $T_c$,
contrary to the phase disordered scenario. A possible
interpretation is that the 
additional induced disorder hinders the percolation
threshold, but does not affect the onset temperature ($T_{onset}$) of
superconductivity in the nanoscale islands.This result confirms
a wide variety of experimental data and theoretical investigations
which have demonstrated that spatially inhomogeneities strongly
affects the HTS properties\cite{Elbio,Yukalov}.

Thus, taking the spatial disorder as an intrinsic phenomena
associated with a PS transition, we reproduce in this paper the
main phase diagram boundaries, the upper and lower pseudogap and
the superconducting phase. We also show that the results of 
Wang et al\cite{Ong} are also compatible with a
disordered material with non-uniform doping level which can take
many  forms, from random patches to stripes. For this purpose, this
paper is organized as follows: In section II, we described briefly
how the CH PS calculations are made (the details are in previous
paper\cite{Otton,js,jpcs}). In section III, we perform the BdG
superconducting calculation on a system which results from the PS.
In section IV, we show the results of both methods combined. In
section V, we generalize a method to calculate the $H_{c2}$ field in
a non-uniform system, again applied to the CH PS results. We finish
with the conclusions.

\section{The Phase Separation }

At least two clearly distinct energy scales are associated with the
pseudogap\cite{Moura,Mello04,Markiewicz} and there are several
indications that the upper pseudogap, which starts at very low
doping and ends near $\rho_m \approx 0.2$\cite{Tallon}, may be a line of
PS (part of this PS line is represented in Fig.(2)). The values of
the upper pseudogap $T^*$, measured by susceptibility, heat
capacity, ARPES, NMR and resistivity, as presented in the review by
Tallon and Loram\cite{Tallon}, or the crossover line shown in Timusk
and Sttat\cite{TS}, seem to be independent of the superconductivity
phase\cite{Tallon}. In fact it was verified recently that this line
falls inside the superconducting dome\cite{Naqib}. The very high
upper pseudogap temperatures in the underdoped region led also Lee
et al\cite{Lee} to argue that it is very unlike any relation to the
superconducting pair formation or to the Nernst onset temperature
$T_{onset}$. As mentioned, a strong evidence towards a spatial PS is
the measurements of $T_1$ over the Cu NQR spectrum of under and
optimally average doping $\rho_m$ samples of bulk LSCO by Singer at
al\cite{Singer}, which was attributed to distributions of local
holes $\rho(i)$ with widths of $\Delta(\rho(i))/\rho_m\approx 0.5$.
They have also showed that the hole segregation increases as the
temperature decreases which may become close to a bimodal with two
branches whose densities are $\rho_m \pm \Delta(\rho(i))$.

In order to harmonize these observations  with theoretical
calculations, we need a model to furnish the local densities in
doped disordered systems. The CH approach is very convenient
because, in general, it describes how the charges tend to a bimodal
distribution below a critical temperature $T_{ps}(\rho_m)$, similar
to what was observed by the NQR measurements\cite{Singer}. Thus we
assume that $T_{ps}(\rho_m)$ is close to the upper pseudogap
line mentioned above\cite{Tallon,Naqib}. As we have discussed
previously\cite{Otton,js,jpcs}, the PS process can lead to different
segregation patterns, depending on the Ginzburg-Landau (GL) free
energy initial coefficients. Since these parameters and the mobility
(which is related to the time scale of the PS process) are not well
known, we adopt values which leads to charge stripes formation, in
order to reproduce the observations in LSCO. The possibilities of
other patterns, like droplets or patchwork was studied
elsewhere\cite{Otton}. Thus, depending on the initial parameters of
the GL free energy, the phase segregation process can form patterns
similar to patchwork\cite{Pan,McElroy} or stripe\cite{Tranquada}
which are observed in HTS and in others high correlated electron
systems\cite{Elbio,Yukalov}. In what follows,  we will take these PS
results as the initial non-uniform input charge distribution on
clusters followed by the BdG local superconducting calculation.

\section{The Local Gap Calculations}

At temperatures below the PS transition, we can observe the
possibility of the local superconducting pairing amplitude
$\Delta(i,T)$ formation at a location ${\bf r}_i$ inside a given
cluster\cite{Franz,Franz2,Ghosal,Edson05,Mello04} as function of the
temperature T. On cooling down the system, the $\Delta(i,T)$ start
at temperatures $T_c(i)$ and increase as the temperature tends to
zero. We find that the $\Delta(i,T)$ have basically the same value
in small regions with the same charge density. This defines what we
call a {\it local superconducting temperature $T_c(i)$} on a site
"i" or in a small cluster. Following the CH results, particularly
those which yield stripe patterns, we have kept fixed the input
local charge densities, which is an approach different than previous
BdG calculations\cite{Franz,Franz2,Ghosal,Edson05}. The details can
be found elsewhere\cite{DDias}, but just for completeness, the BdG
equations are,
    \be
    \begin{pmatrix}
    \xi & \Delta \\
    \Delta^* & -\xi^*
    \end{pmatrix}
    \begin{pmatrix}
u_n({\bf r}_i) \\
v_n({\bf r}_i)
\end{pmatrix}
    =
    E_n
    \begin{pmatrix}
u_n({\bf r}_i) \\
v_n({\bf r}_i)
\end{pmatrix},
    \lb{BdG}
    \ee
where $E_n \ge 0$ are the quasiparticles energies, and

    \beqn \nb
    \xi u_n({\bf r}_i)&=&-\sum_{\delta}t_{i,i+\delta}u_n({\bf r}_i+{\bf\delta})-\widetilde{\mu}_iu_n({\bf r}_i),\\
    \Delta u_n({\bf r}_i)&=&\Delta_U({\bf r}_i)+\sum_{\delta}\Delta_{\delta}({\bf r}_i)u_n({\bf
    r}_i+{\bf\delta}),
    \eqn
the t's are the hopping parameters and $\widetilde{\mu}_i$ the local
chemical potential. There are also similar equations for $v_n({\bf
r}_i)$. When these BdG equations are solved, they give the
eigen-energies $E_n$ and the local amplitudes $u_n({\bf r}_i)$ and
$v_n({\bf r}_i)$. Through these amplitudes and the positive
eigen-energies it is possible to calculate the pairing amplitudes
    \be
\Delta_U({\bf r}_i)=
    -U\sum_nu_n({\bf r}_i)v^*_n({\bf
    r}_i)\tanh\dfrac{E_n}{2K_BT},\lb{D0local}\ee
    \beqn \nb
    \Delta_{\delta}({\bf r}_i)&=&-\frac{V}{2}\sum_n[u_n({\bf r}_i)v^*_n({\bf
    r}_i+{\bf\delta})\\&&+
    v^*_n({\bf r}_i)u_n({\bf
    r}_i+{\bf\delta})]\tanh\frac{E_n}{2K_BT}
\lb{Delta}
    \eqn
and the local hole density is given by
    \be \rho({\bf r}_i)=1-2\sum_n[\vert u_n({\bf r}_i)\vert^2f_n+\vert v_n({\bf
    r}_i)\vert^2(1-f_n)],
\lb{rolocal}
    \ee
where $f_n$ is the Fermi function. The BdG equations (\ref{BdG}) are
solved self-consistently together with the equations for the pairing
amplitudes (Eq.(\ref{Delta})) and for the local hole
density(Eq.(\ref{rolocal})) which is kept fixed throughout the
entire calculation. For calculations using d-wave symmetry we have
$V<0$ and $U>0$ and Eq.(\ref{Delta}) can be written as\cite{Franz2}
    \be \Delta_d({\bf r}_i)=
    \frac{1}{4}[\Delta_{\widehat{x}}({\bf r}_i)+\Delta_{-\widehat{x}}({\bf r}_i)
    -\Delta_{\widehat{y}}({\bf r}_i)-\Delta_{-\widehat{y}}({\bf
    r}_i)].
    \ee

In the calculations we have used the hopping parameters up to third
neighbors,  close to those derived from the ARPES data for
YBCO\cite{Schabel}, that is, $t=0.23eV$ (first neighbors),
$t_2=-0.61t$ and $t_3=0.2t$. These values are slightly different
from our other work\cite{DDias}, which we used hopping values up to
$5^{th}$ neighbors following the ARPES results\cite{Schabel}. In fact
we made various studies around the ARPES data, and they all give the
same qualitative results, and therefore, we present the calculations
that yielded the best quantitative agreement with the experiments.
The potentials are $U=1.1t$ and $V=-0.6t$ which are also close to
previous calculations\cite{Edson05}. The value of $t=0.23eV$ was
chosen in order to reproduce\cite{DDias} the measurements of the
zero temperature superconducting gap by Harris et al\cite{Harris}
and the ARPES leading edge shift by Ino at al\cite{Ino}. Furthermore
this value of $t$ is in the range of several experiments on
HTS\cite{scbook}.

\section{Results}

    \begin{figure}[!ht]
    \includegraphics[width=9cm]{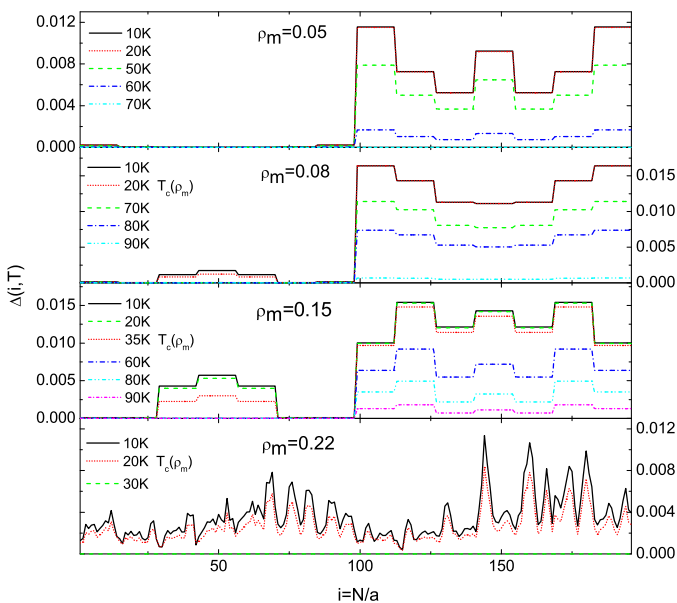}
    \caption{(color online)Temperature evolution of the local pairing amplitude $\Delta(i,T)$,
in units of $t=0.23eV$ at each site "i" of a $14\times14$ cluster
with 196 sites. Because the CH phase separation for $\rho_m\le0.20$
and the stripe pattern, the sites in the left have $\rho(i)\approx
0$ and the ones in the right $\rho(i)\approx 2\rho_m$. In contrast,
$\rho_m=0.22$ which is beyond the phase separation limit $T_{ps}$,
has just a Gaussian charge disorder and its profile is very
different.} \label{sitios1}
    \end{figure}

Let큦 apply our method to the LSCO system. We then perform
calculations with stripe disordered clusters as derived from a set
of parameters in the CH approach and keep this initial doping
configuration fixed in the temperature range of our calculations. As
we have already mentioned, the $T_{ps}(\rho_m)$ ends near $\rho_m\ge
0.20$\cite{Tallon}, and for higher density values there is not any
PS, only small fluctuations around $\rho_m$. Fig.(\ref{sitios1})
shows the temperature evolution of the superconducting local pair
amplitudes $\Delta({\bf r_i},T)$  or simply $\Delta(i,T)$ at each
site $i$ in a square mesh $14\times14$.

Following the charge patterns derived from the CH calculations, the
7 stripes at the left are characterized by local doping
$\rho(i)\approx 0$ and the 7 at the right side have $\rho(i)\approx
2\rho_m$. This is the scheme of the three upper panel 
of Fig.(\ref{sitios1}). The high values of the
$T^*\approx T_{ps}$ implies that lightly doped compounds like the
$\rho_m=0.05$ has $\rho(i)$ strictly zero in the low doping region.
At low temperatures, the superconducting regions for this compound,
those with a finite value of $\Delta(i,T)$, develops only in the
high density region and all together, they never reach more than
$50\%$ of the total sites. Consequently, the superconducting sites
never percolate and there is no superconducting phase for this
compound. For doping of $\rho_m > 0.06$ the low doping sites have
some residual fluctuation(of the phase separation process) $\rho(i)>
0$ which grows with $\rho_m$, from $ \approx 0.03$ to $0.05$, what
changes the properties of a sample from a disordered insulator  into
a disordered metal, with a superconducting phase at low
temperatures. Thus, for compounds with $0.06<\rho_m<0.20$, at low
temperatures, one can see in Fig.(\ref{sitios1}), that $\Delta(i,T)$
develops also at the very low doping regions, i.e., in the left
region of the clusters represented in Fig.(\ref{sitios1}). At the
temperature $T_c(\rho_m)$, when the $\Delta(i,T)$ arise at these low
doping sites, the system becomes superconductor either by the
percolation of the many local superconducting regions (the regions
where $\Delta(i,T)$ are non-vanishing)  or by Josephson
coupling\cite{OWK,Mello03}. The values of $T_c(\rho_m)$ are shown in
each panel of Fig.(\ref{sitios1}). Consequently, below
$T_c(\rho_m)$, superconducting critical temperature, the system can
hold a dissipation-less current, assuming, as usual in mean field
(BCS) theories, that all these superconducting regions form a
conventional superconducting phase with a {\it unique  phase
$\theta$} which is essential to percolation and Josephson
coupling. Above $T_c(\rho_m)$ the compounds form a mixture of
superconducting, insulator and normal domains and above the pairing
formation temperature $T_{onset}(\rho_m)$, they are disordered
metals with mixtures of normal ($\rho(i) \ge 0.03-0.05$) and
insulator ($\rho(i)\le 0.03-0.05$) regions.

    \begin{figure}[!ht]
    \includegraphics[width=9cm]{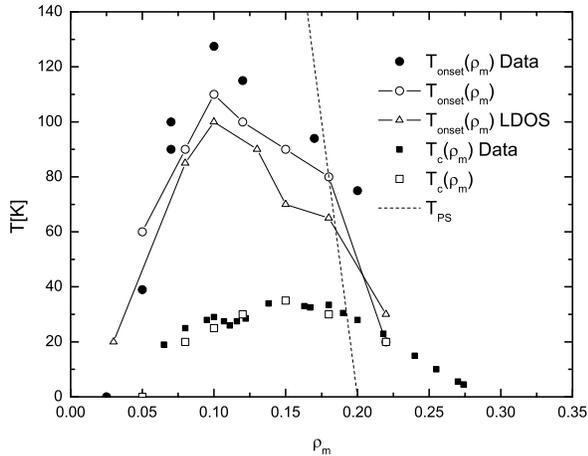}
    \caption{The local superconducting onset temperature taken as
the Nernst signal (solid line, from Ref.14) and the superconducting percolation
temperature $T_c(\rho_m)$. The  black circles and squares are the
    experimental data of $T_{onset}$ and $T_c$, respectively. The triangles
are the values derived from the density of states. The
phase separation line $T_{ps}$ is also shown.}
\label{PhDiag}
    \end{figure}

From these results, we identify $T_{onset}(\rho_m)$ as the highest
temperature to induce a $\Delta(i,T)$ in any region of a given
compound which is easily seen from the panels of Fig.(\ref{sitios1})
(and from similar studies on others values of $\rho_m$). Thus
$T_{onset}(\rho_m)$ is identified with the onset of Nernst signal
because the rising of superconducting regions in a metallic matrix
increases the vortices drift. The values of $T_{onset}(\rho_m)$ and
$T_c(\rho_m)$ are shown in Fig.(\ref{PhDiag}).  The maximum pairing
amplitudes for each $\rho_m$ at low temperature ($\Delta_0(p_m)$) is
in good agreement with the ARPES zero temperature leading edge
shift\cite{DDias} or the maximum magnitude of the superconducting
gap\cite{Harris,Ino}.

\begin{figure}[!ht]
    \includegraphics[width=9cm]{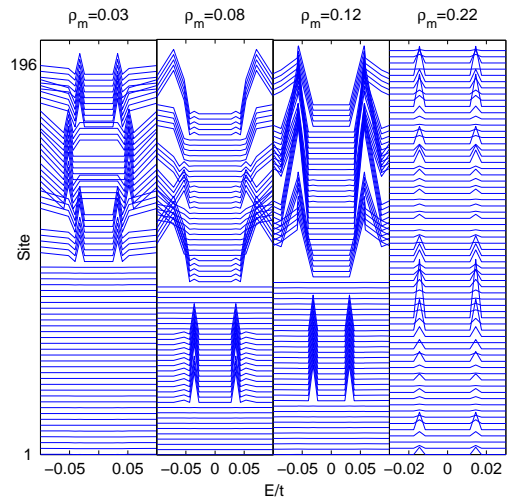}
    \caption{The study of the superconducting gap from 
 the first peak of the local density of states (LDOS)
at zero temperature. At $\rho_m=0.8$ the superconducting gap is of
the order of 20 meV and decreases as $\rho_m$ increases, which
agrees with the measurements made by the ARPES\cite{Harris,Ino}}
\label{DE}
\end{figure}

The zero temperature gap can be also obtained by a different
procedure through the study of the local density of states (LDOS),
which is given by $N_i(E)=\sum_n[|u_n({\bf
x}_i)|^2f_n^{'}(E-E_n)+|v_n({\bf x}_i)|^2f_n^{'}(E+E_n)]$, where
the prime is the derivative with respect to the argument. In a
typical cluster of our calculations,
the opening of this maximum $\Delta(i,T)$, occurs at the  neighbor of site
$i=127$ (see, for instance, the compound with $\rho_m=0.15$ in Fig.(\ref{sitios1}))
and $T_{onset}(\rho_m)$ was defined as the temperature which such
local superconducting gap vanishes. 
Now, by the local density of states, we can analyze the local gaps and take the superconducting
gap as the minimum value of $E_n$ with a
non-vanishing spectral weight, as shown in Fig.(\ref{DE}). 
The corresponding temperature which this superconducting gap
vanishes is also shown in Fig.(\ref{PhDiag}) and it
is quite similar to $T_{onset}(\rho_m)$, in fact, we believe they
will be equal for larger clusters.  Thus, both ways
of defining the onset temperature of superconducting gap of the system are in good
agreement with the Nernst signal onset temperature.

\section{The Critical Field}
The presence of superconducting regions with different critical
temperature will affect the magnetic response. Furthermore, those
regions with the local critical temperature $T_c(i)$ above the
superconducting critical temperature $T_c(\rho_m)$ greatly affects
the normal state properties. Such effects have been seen by many
measurements of anomalous magnetic properties of the normal
phase\cite{Edson,Riga,JL,Cabo}. In particular, the anomalous upper
critical field $H_{c2}$ was recently measured by Wang et
al\cite{Ong} on the LSCO system.

A few years ago some of us have developed a way to
calculate $H_{c2}$ of a disordered superconductor\cite{Edson}. While
in that paper we have dealt with a general disorder, here we apply
this method to the stripe disorder described above, appropriated to
the LSCO system. To explain it, we follow along the lines described
by Caixeiro et al\cite{Edson}. The GL upper critical field of a
homogeneous superconductor may be written as\cite{Edson}

\begin{eqnarray}
H_{c2}(T)&=&{\Phi_{0}\over 2\pi\xi_{ab}^2(0)}\left( {T_c-T\over T_c}\right).\hspace{0.5cm} (T<T_c)
\label{3}
\end{eqnarray}

\begin{figure}[!ht]
\includegraphics[width=9cm]{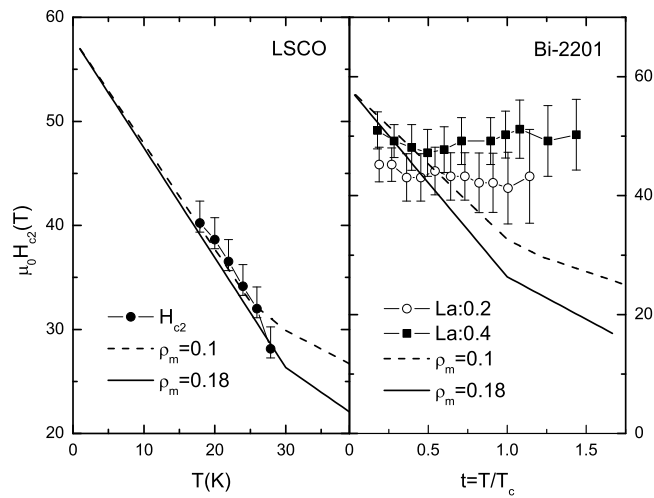}\\
\caption{The lines are calculations of $H_{c2}$ considering samples
with different local superconducting temperatures $T_c(i)$, similar
to Fig.(\ref{sitios1}). Left panel are comparison with the
experimental points from Ref.(\cite{Nernst3}) for LSCO
($\rho_m=0.2$) and the right panel with points from
Ref.(\cite{Ong}).} \label{Hc2}
\end{figure}

Let us now apply this expression to a HTS with stripes
inhomogeneities in the charge distribution. As showed in
Fig.(\ref{sitios1}) , when a superconducting amplitude develops in a
given  region ``${\bf r_i}$'', it has a local superconducting
temperature $T_c(i)$ and it will contribute to the  critical field
with a local linear upper critical field $H_{c2}^i(T)$ below
$T_c(i)$. Therefore, the total contribution of the various local
superconducting regions to the upper critical field is the sum of
all the $H_{c2}^i(T)$'s at temperatures below $T_{onset}(\rho_m)$,
which is the temperatures that the system starts to develop some
superconducting region. In this way, the $H_{c2}$ for a whole sample
is

\begin{eqnarray}
&&H_{c2}(T)={\Phi_{0}\over 2\pi\xi_{ab}^2(0)}{1\over W}\sum_{i=1}^W
\left( {T_c(i)-T\over T_c(i)}\right)\nonumber \\
&&={1\over W}\sum_{i=1}^W H_{c2}^i(T)\hspace{0.25cm} (T<T_c(i)\le
T_{onset}(\rho_m)) \label{Hc2eq}
\end{eqnarray}
where $W$ is the number of superconducting regions, (the high hole
densities stripes), each with its local $T_c(i)$$\le
T_{onset}(\rho_m)$. For the LSCO series a coherence length of
$\xi_{ab}(0)\approx 22\AA$ is in agreement with the
measurements\cite{Edson}. This value of $\xi_{ab}(0)$ leads to
$H_{c2}(0)$=$\Phi/2\pi\xi_{ab}^2(0)$=64T. This value is close with
the extrapolated values as it is seen in Fig.(\ref{Hc2}).

In Fig.(\ref{Hc2}) we plot the results for $\rho_m=0.1$ and $0.18$
and compared with the experimental values of Wang et al\cite{Ong}
for overdoped Bi2201(La:04 and 02). We also compare our calculations
to the previous experimental values from the same
group(\cite{Nernst3}) on LSCO because our calculations are directed
to reproduce the non-uniformities on this series. The excellent
agreement indicates that LSCO compounds have indeed a distribution
of local superconducting regions $T_c(i)$ close to that derived
here(see Fig.(\ref{sitios1})). On the other hand, since the $H_{c2}$
curves of Bi2201 are so flat near $T/Tc\approx 1$, according our
results, it indicates that this material is highly disordered in
$T_c(i)$, with $T_{onset}(\rho_m)$ much larger than critical
temperature $T_c(\rho_m)$.

A similar procedure, taken into account a non-uniform system with
variations on the local $T_c(i)$, was applied before to reproduce
experimental values of the magnetization above $T_c$ on LSCO single
crystals\cite{Riga}, and the same results\cite{JL} are in good
qualitative agreement with the measurements of Wang et al\cite{Ong}.
Recently, a distribution of local $T_c(i)$ was also used to
reproduce and to interpret the magnetization data above
$T_c$\cite{Cabo}.

\section{Conclusion}

Taking the upper pseudogap line as a PS transition temperature, we
have monitored the development of charge inhomogeneities by the CH
theory which describes how charge segregation increases as the
temperature goes down, similarly as seen in the NQR experiments\cite{Singer}. 
In this way we have followed the process of charge segregation in the
LSCO system and studied the superconducting properties of the normal
and superconducting phases as function of doping.  Our approach is
general and could be directed to other patterns of charge disorder.
Each charge domain with constant density, in general, is
characterized by its  local superconducting pairing amplitude
$\Delta(i,T)$ which arises at $T_c(i)$.  This introduces the new 
concept of a local superconducting
temperature, which marks the appearence of the local pairing amplitude . The
maximum local value of $T_c(i)$ of a given compound with average
doping level $\rho_m$, as we
demonstrated, can be related to the Nernst onset temperature
$T_{onset}(\rho_m)$ or lower pseudogap. This approach, with different
regions or islands with local superconducting tempeartures $T_c(i)$ provides a
clear explanation to the non-vanishing of  $H_{c2}$ above the critical
temperature $T_c(\rho_m)$. It also explains why early tunneling experiments\cite{Moura,Harris}
did not see any special signal at $T_c(\rho_m)$.

In conclusion, we have provided the steps to an interpretation of the
HTS phase diagram where the disorder plays a key role. We have 
also shown that some physical properties, like the experimental data of Wang et
al\cite{Ong} on the Nernst effect and $H_{c2}$, which were presented
as evidences of the phase disordered scenario, can also be favorable
interpreted as special features of conventional superconductivity
which develops in spatially disordered systems at 
higher hole densities regions. Our calculations furnish also an
interpretation to the normal phase as a disordered metal and to the
three most measured phase diagram boundaries in HTS: the upper
pseudogap as a phase separation transition, the lower pseudogap as
the onset of islands of superconductivity and the system superconducting
critical temperature as the Josephson coupling or percolation
temperature among the various superconducting regions.

This work has been partially supported by CAPES, CNPq and CNPq-Faperj
Pronex E-26/171.168/2003.


%

\end{document}